# Possible quantum criticality and coexistence of spontaneous ferromagnetism and field-induced metamagnetism in triple-layered $Sr_4Ru_3O_{10}$

G. Cao, S. Chikara and J. W. Brill

Department of Physics and Astronomy, University of Kentucky, Lexington, KY 40506

P. Schlottmann

Department of Physics, Florida State University, Tallahassee, FL 32306

Abstract

A thermodynamic and transport study of $Sr_4Ru_3O_{10}$ as a function of temperature and magnetic field are presented. The central results include a growing specific heat C with increasing field B, a magnetic contribution to C/T at low temperatures proportional to –log(T), an abrupt jump and a peak in C/T at 2.90 T and 7 T for B||ab-plane and B||c-axis, respectively, and corresponding changes in the low T power laws of the resistivity. The novelty of this work lies in the fact that the quantum fluctuations occur in a system that shows both intralayer metamagnetism and interlayer *spontaneous* ferromagnetism, a feature characteristically different from all other systems involving quantum criticality.

A quantum critical point (QCP) can be realized by tuning the critical temperature of a phase transition to absolute zero by varying external parameters such as the magnetic field, pressure or the doping level. While a phase transition at finite T is governed by thermal fluctuations, the relevant fluctuations close to a QCP are of quantum mechanical nature. A QCP usually impacts the physical properties over a wide range of temperatures. Quantum criticality is well illustrated in heavy fermions [1] and high $T_c$ cuprates, and, recently, in materials intimately associated with itinerant metamagnetism. For instance, $UPt_3$ [2], $CeRu_2Si_2$ [3] and $Sr_3Ru_2O_7$ [4] are enhanced paramagnets showing field-induced ferromagnetism via a first-order metamagnetic transition. When the critical end point of the first order metamagnetic transition is tuned to zero temperature, the critical fluctuations cause the breakdown of Fermi liquid behavior with profound consequences, such as divergent specific heat [C/T ~ -log(T)] and unusual power laws in the resistivity at low T [1-6]. In this Letter, we report the temperature and field dependence of the specific heat, resistivity and magnetization of triple-layered $Sr_4Ru_3O_{10}$. The results reveal a host of anomalous properties, namely, a growing specific heat C with increasing magnetic field B, a –log(T) contribution to C/T at low temperatures, an abrupt jump and anomaly in C/T at $B_c$=2.90 T and $B_c$=7 T for B||ab-plane and B||c-axis, respectively, and corresponding changes in the power law of the resistivity between $T^2$ and $T^{3/2}$. The novelty of this work lies in the fact that the quantum fluctuations occur in a system where both interlayer spontaneous ferromagnetism and intralayer field-induced metamagnetism coexist, a feature distinctively different from all other systems involving a QCP.

$Sr_4Ru_3O_{10}$ belongs to the layered Ruddlesden-Popper series, $Sr_{n+1}Ru_nO_{3n+1}$ (n=number of Ru-O layers/unit cell). The comparable and competing energies of crystalline fields (CEF), Hund's rule interactions, spin-orbit coupling and electron-lattice coupling crucially determine the CEF level splitting and the band structure, hence the ground state. The physical properties are thus highly dimensionality (or n) dependent. As a result, the magnetic state of $Sr_{n+1}Ru_nO_{3n+1}$ systematically evolves from paramagnetism (n=1), enhanced paramagnetism (n=2) to spontaneous ferromagnetism (n=∞) with increasing n. Situated between n=2 and n=∞, $Sr_4Ru_3O_{10}$ (n=3) displays interesting phenomena ranging from quantum oscillations [8], tunneling magnetoresistance [8], unusual low temperature specific heat [9], strong spin-lattice coupling [10, 11] to switching behavior [12]. The unique feature, however, is borderline magnetism: For a field *along the c-axis* (perpendicular to the layers), $Sr_4Ru_3O_{10}$ displays spontaneous ferromagnetism, while for a field within *the ab-plane* it features a pronounced peak in magnetization and a first-order metamagnetic transition [7-11]. The ferromagnetism along the c-axis indicates that the Stoner criterion is satisfied, $Ug(E_F) \geq 1$, where U is an exchange interaction and $g(E_F)$ the density of states at the Fermi surface. Below $T_C$ the spin-up and spin-down bands are thus spontaneously split by the exchange slitting Δ in the absence of an applied magnetic field. The response to a field in the plane, however, is strikingly similar to Stoner enhanced paramagnetism and metamagnetism with $Ug(E_F)<1$ [4,13,14]. The instabilities and the anisotropy may arise from the two-dimensional Van Hove singularity (logarithmical divergence) close to the Fermi level [15] in conjunction with the coupling of the spins to the orbital states of $Ru^{4+}$. The coexistence of the interlayer ferromagnetism and the intralayer metamagnetism makes $Sr_4Ru_3O_{10}$ crucially different from $Sr_3Ru_2O_7$ where spontaneous ferromagnetism is absent.

The properties of $Sr_4Ru_3O_{10}$ critical to the discussion are the following: (1) The c-axis magnetization, $M_C$, is ferromagnetic with $T_C$ at 105 K followed by an increased spin polarization below $T_M$=60 K with large irreversibility upon in-field and zero-field cooling. In contrast, the ab-plane magnetization, $M_{ab}$, is much smaller; it exhibits a weak cusp at $T_C$ and a broad peak at $T_M$, resembling enhanced paramagnetic behavior [8]. (2) The isothermal magnetization $M_C$ illustrates that at low T the spins are readily polarized and saturated along the c-axis at B=0.2 T, yielding a saturation moment $M_s$ of 1.2 $\mu_B$/Ru extrapolated to B=0 for T=1.7 K. On the contrary, $M_{ab}$ displays a first-order metamagnetic transition at $B_c$, which disappears for T larger than $T_M$ [8]. (3) The temperature dependent (B=0) specific heat of the present sample is the same as that we previously reported for a sample measured with ac-calorimetry [9]. In particular, there is a mean-field like anomaly $\Delta C \sim R/3$, where R = 8.31 J $mol^{-1}$ $K^{-1}$ is the gas constant, at Tc, but no anomaly is observed at $T_M$, suggesting a cross-over rather than a phase transition at this temperature [9]. At low temperature, there is anomalous negative curvature in a conventional C/T vs. $T^2$ graph, which will be discussed further below.

The single crystals studied were grown using flux techniques [16] and characterized by single crystal x-ray diffraction at 90 K and room temperature, EDX and TEM. All results suggest that the crystals studied are of high quality with *no impurity phases and no intergrowth*. The high-quality of our samples is further confirmed by the observation of quantum oscillations and a small Dingle temperature of 3 K, a measure of impurity scattering [8]. Heat capacity measurements were performed using a Quantum Design (QD) PPMS which utilizes a thermal-relaxation calorimeter operating between 1.8 and 400 K in fields up to 9 T. Magnetic and transport properties were measured using a QD magnetometer (7T XL) and a 15 T Oxford magnet along with a Linear Research 700 ac Resistance Bridge.

The application of B tends to align spins and usually suppresses spin fluctuations and hence the specific heat. It is striking that the specific heat of $Sr_4Ru_3O_{10}$ responds to B oppositely. Fig.1 shows the specific heat divided by temperature as a function of T for 1.8≤T≤12 K with B||ab-plane (a) and c-axis (b), respectively. Data for T>12 K is not shown for clarity. There are a few crucial features. Firstly, the low temperature C/T increases radically for B||ab-plane, particularly in the vicinity of the metamagnetic transition (Fig.1a) in contradiction to the anticipated behavior. For example, C/T at 1.8 K jumps by 20 mJ/mol $K^2$ in just a 0.05 T interval from 2.90 and 2.95 T, implying a considerable enhancement of the quasi-particle m* in the density of states $g(E_F)$. Secondly, C/T for B||c-axis shows vastly different T-dependence (Fig.1b), decreasing with increasing B for T>6 K. This is consistent with a suppression of spin fluctuations as anticipated for a regular magnetic state. But it grows, though less drastically, for T<6 K by showing a broad maximum near 8 T, suggesting that unexpected low-energy excitations develop in the spontaneous ferromagnetic state. Thirdly, C/T for B≤2.7 T and T < 10 K nearly follows a linear T-dependence, C/T~*a*+*b*T. The rapid increase of *a* with B could be indicative of a Fermi liquid with a nearby 2D critical point [6]. This interpretation requires a very small phonon contribution, $\beta T^2$ (apparent in the positive curvature in C/T vs. T for T > 10 K) and therefore an unphysically large value of the Debye temperature. Alternatively, as discussed in [9], the specific heat can be fit by $C/T = \gamma + \beta T^2 + \delta T^{1/2}$ (with a typical value of $\beta \sim 0.04$ mJ $mol^{-1} K^{-4}$), where the $T^{1/2}$ term is associated with an unexpectedly large contribution from ferromagnetic spin-waves. As B is increased further the T-dependence of C/T weakens from almost T-linear at $B<B_c$ to a shoulder with a small peak at B=5 T and a plateau for B≥6T (Fig.1). To further emphasize the unusual temperature dependence of C/T in fields, C/T vs. $T^2$ is plotted in Fig.1c, where the negative curvature associated with the $T^{1/2}$ term is manifested for T< 8 K [9]. The inset clearly shows the sharp peak at 2.5 K for B=5 T that diminishes at higher fields and eventually evolves into a rapid downturn at 9 T.

It is clear that the amplitude of fluctuations rapidly grows as B increases. To separate the field-induced contribution ΔC from other contributions to C, we subtract the zero-field C(0) from the in-field C(B), i.e., ΔC=C(B)-C(0). ΔC/T plotted as a function of T is shown in Fig.2 for B||ab-plane (Fig.2a) and B||c-axis (Fig.2b). A dominant feature is that ΔC/T increases logarithmically with decreasing T as B rises. The slope of the log-dependence gets stronger when B approaches 5 T and 7 T for B||ab-plane and B||c-axis, respectively. At 1.8 K, ΔC/T for B||ab-plane increased by a factor of three from less than 0.04 J/mol $K^2$ at 1 T to nearly 0.12 J/mol $K^2$ at 5 T. At higher temperature, there exists a broad peak marked by an arrow. It moves to lower temperatures and eventually vanishes as B increases so that the entropy is shifted into the logarithmic upturn. Remarkably, for B||ab-plane, a pronounced peak develops for B=2.9 T, which then broadens for $B>B_c$ (Fig.2a), whereas for B||c-axis, the peak is progressively suppressed by B and vanishes at B>6 T. The emergence of the peak is a signature of the large spin fluctuations in the vicinity of a quantum phase transition.

The abrupt jump in C/T and the hysteresis at the metamagnetic transition for B||ab-plane are once more emphasized in Fig.3, in which the field dependence of C/T is compared to that of the resistance and magnetization. The jump in C/T is drastic and

persists up to ~12 K, suggesting a critical end point of the metamagnetic transition (Fig.3b). As shown in Fig.3c (left scale) the metamagnetic transition also affects the magnetoresistence ratio (current along the c-axis), defined as $\Delta\rho_c/\rho_c(0)=[\rho_c(B)-\rho_c(0)]/\rho_c(0)$, which changes by more than 40% near $B_c$, confirming large spin fluctuations in a state without long-range order immediately above the transition. The anomaly at 8 T for B||c-axis is clearly seen in C/T (Fig.3a) and as a kink in $\Delta\rho_c/\rho_c(0)$ (Fig.3c).

The presence of the quantum fluctuations is further corroborated by the temperature dependence of the resistivity. Shown in Fig.4a is the ab-plane resistivity $\rho_{ab}$ at a few representative fields as a function of $T^2$ or $T^{5/3}$ (upper axis) for a range of 1.7<T<17 K. The linearity of $\rho_{ab}$ in the plot suggests well-defined power laws followed at various B. Fig.4b maps the details of the coefficients of the $T^2$- and $T^{5/3}$-dependences of $\rho_{ab}$ as a function of B. $\rho_{ab}$ fits well to $\rho_{ab}=\rho_o+A_{ab}T^2$ for B<2 T where $\rho_o$, the residual resistivity, is 6 μΩ cm at B=0, indicative of the high purity of the sample. The coefficient $A_{ab}$, which depends on the effective mass m*, rapidly increases with B, indicating a vanishing Fermi temperature and a divergent m*. $\rho_{ab}$ starts to deviate from the $T^2$-dependence at B=2.2 T, signaling the breakdown of Fermi liquid properties. In the region of $2.2\leq B<5.5$ T $\rho_{ab}$ is proportional to $A_{ab}{}^*T^\alpha$, where $\alpha$ is smaller than 2. $\alpha$ briefly varies between 1.5 and 1.6 for 2.0<B<2.4 T and then settles at 5/3 for $2.4\leq B\leq 5.5$T. The coefficient $A_{ab}{}^*$ for the $\alpha=5/3$ dependence rises steeply with B, peaks at 2.9 T and decreases for larger fields. The Fermi liquid behavior is recovered for B>5.5 T. This is consistent with the behavior of C which at low T starts to decrease when B>5 T (see Fig.1a). The residual resistivity, $\rho_o$, consistently shows a similar field dependence and diverges near $B_c$ as reported earlier [8].

A singular T-dependence of the resistivity with $\alpha < 2$, specifically, $\alpha=3/2$ and 5/3 is seen in systems with quantum criticality such as MnSi [13], $Sr_3Ru_2O_7$ [4], heavy fermion systems [5] as well as impurity doped $Sr_4Ru_3O_{10}$ [11]. The $T^{5/3}$-dependence of resistivity is often attributed to either dominating low-angle electron scattering (low-*q* fluctuations) or high-*q* fluctuations scattering electrons in the vicinity of a critical point [13,16-18], hence weakening the temperature dependence of the resistivity from $T^2$. The power-law $T^{3/2}$ is thought to be associated with effects of diffusive motion of the electrons caused by the interactions between the itinerant electrons and critically damped magnons [13]. The change of $A_{ab}$ and $A_{ab}$* with B entirely tracks C/T and $M_{ab}$, pointing to the proximity of a critical point and an intimate connection between the critical fluctuations and the metamagnetism

What is equally intriguing is that C/T for B||c-axis, where the spontaneous ferromagnestism occurs, shows a weaker and yet well-defined peak at 7 T followed by a minimum at $B_m$(c)=8 T (Fig.3a). Unlike the jump in C/T for B||ab-plane, this peak sensitively changes with T, and nearly vanishes for T>6.8 K, where C/T no longer increases with B, the behavior expected for a regular metal. The anomaly in C/T also accompanies a varying $\alpha$ of resistivity between 3/2 and 5/3 near 7 T for B||c-axis (not shown), and an abrupt change in slope of $\rho_c$ at $B_m$(c)=8 T (Fig.3c, left scale), suggesting a significant change in the scattering rate. This behavior brings up an interesting question: Why do the spontaneously spin-split bands undergo the critical fluctuations? It is established that the metamagnetic critical fluctuations essentially arise from field-induced spin-split Fermi surfaces that are not well defined and/or Bragg diffraction of electrons off a quantum critical spin density wave. Therefore $g(E_F)$ or m* and long range

correlations diverge at the critical point, leading to non-Fermi liquid behavior. The c-axis ferromagnetic state in $Sr_4Ru_3O_{10}$ (see $M_c(B)$ in Fig.3c) is robust [7-10], thus the spontaneously spin-split bands along with $g(E_F)$ are stable and well defined. Therefore field-driven quantum critical fluctuations and nesting properties are entirely unanticipated.

Long-range ferromagnetic order and a metamagnetic transition have been predicted within a simple Stoner model with a two-dimensional density of states, which has a logarithmically divergent Van Hove singularity [15], but this model is too simple to explain the magnetic anisotropy, which requires a coupling of the spins to the orbits. The $t_{2g}$-orbitals are split in the band structure due to the crystalline electric fields, so the xy-orbital (contained in the ab-plane) is favored. As a consequence of this splitting orbital order is expected. Together with the Hund's rule couplings (maximizing the spin and the spin-obit coupling) the spin and the orbital states are coupled perturbatively, favoring the c-axis ferromagnetism and the ab-plane metamagnetism. This work highlights the complexity of such couplings and provides strong evidence for the existence of quantum critical fluctuations driven by both spontaneous ferromagnetism and field-induced metamagnetism. These results, which cannot be accounted for using currently existing models, strongly suggest an exotic ground state.

**Acknowledgements**: G.C. is thankful to Prof. Lance DeLong for useful discussions. This work was supported by NSF grants DMR-0240813, DMR-0400938, DMR-0552267 and DOE grant DE-FG02-98ER45707.

**References:**

1. G.R. Stewart, Rev. Mod. Phys. **73**, 797 (2001)
2. J.S. Kim, D. Hall, K. Heuser, G. R. Stewart, Solid State Commun. **114** 413 (2000)
3. S. Kamble, H. Suderow, J. Flouquet, P. Haen, P. Lejay, Solid State Commun. **95**, 449 (1995); Y. Aoki, et al., J. Magn. Magn. Mater. **177-181** 271 (1998)
4. S.I. Ikeda, Y. Maeno, S. Nakatsuji, M. Kosaka and Y. Uwatoko, Phys. Rev. B **62** R6089 (2000); R.P. Perry, L.M. Galvin, S.A. Grigera, A.J. Schofield, A.P. Mackenzie, M. Chiao, S.R. Julian, S.I. Ikeda, S. Nakatsuji, and Y. Maeno, Phys. Rev. Lett. **86**, 2661 (2001); S.A. Grigera, R.P. Perry, A.J. Schofield, M. Chiao, S.R. Julian, G.G. Lonzarich, S.I. Ikeda, Y. Maeno, A.J. Millis and A.P. Mackenzie, Science **294**, 329 (2001); Z.X. Zhou, S. McCall, C.S. Alexander, J.E. Crow, P. Schlottmann, A. Bianchi,, C. Capan, R. Movshovich, K. H. Kim, M. Jaime, N. Harrison, M.K. Haas, R.J. Cava and G. Cao, Phys. Rev. B **69** 140409(R) (2004)
5. P. Coleman and C. Pepin, Physica B **312-313** 383 (2002)
6. A.J. Millis, A. J. Schofield, G.G. Lonzarich, and S.A. Grigera, Phys. Rev Lett. **88**, 217204 (2002)
7. M. Crawford, R.L. Harlow, W. Marshall, Z. Li, G. Cao, R.L. Lindstrom, Q. Huang, and J.W. Lynn, Phys. Rev B **65**, 214412 (2002)
8. G. Cao, L. Balicas, W.H. Song, Y.P. Sun, Y. Xin, V.A. Bondarenko, J.W. Brill, S. Parkin, and X.N. Lin, Phys. Rev. B **68**, 174409 (2003)
9. X.N. Lin, V.A. Bondarenko, G. Cao, J.W. Brill, Solid State Comm. **130**, 151 (2004)

10. B. Gupta, M. Kim, H. Barath, S.L. Cooper and G. Cao, Phys. Rev. Lett, **96,** 067004 (2006)

11. S. Chikara, V. Durairaj, W.H. Song, Y.P. Sun, X.N. Lin, A. Douglass, G. Cao and P. Schlottmann, Phys. Rev. B, 2006, in print

12. Z. Mao, et al. cond-mat/0406439.

13. C. Pfleiderer. S.R. Julian and G. G. Lonzarich, Nature **414**, 427 (2001)

14. T. Sakakibara, T. Goto, K. Yoshimura, and K. Fukamichi, J. Phys.: Condensed Matter **2**, 3381 (1990)

15. B. Binz and M. Sigrist, Europhys. Lett. **65**, 816 (2004)

16. T. Moriya, *Spin Fluctuations in Itinerant Electron Magnetism*, (Springer-Verlag, Berlin) (1985)

17. A.J. Millis, Phys. Rev. B **48**, 7183 (1993)

18. C. Pfleiderer, G.J. McMullan, S.R. Julian and G.G. Lonzarich, Phys. Rev. B **55**, 8330 (1993)

**Figure Captions:**

**Fig.1**. Specific heat divided by temperature, C/T, as a function of T for (a) B||ab-plane and (b) B||c-axis; (c) C/T for B||ab-plane as a function of $T^2$ for 0≤B≤5 T; Inset: C/T for B||ab-plane vs. $T^2$ for 5≤B≤9 T.

**Fig.2**. Logarithmic temperature dependence of [C(B)-C(0)]/T for (a) B||ab-plane and 0≤B≤5 T (inset: for 5≤B≤9 T), and for (b)B||c-axis. The arrows track the shift of the peak.

**Fig.3**. Field dependence of C/T for (a) B||c-axis and (b) B||ab-plane for various temperatures; (c) [ρ(B)-ρ(0)]/ρ(0) (left scale) at T=0.6 K and isothermal magnetization M for B||ab-plane and B||c-axis (right scale) at T=1.7 K.

**Fig.4.** (a) The ab-plane resistivity $\rho_{ab}$ at a few representative fields as a function of $T^2$ (lower axis) or $T^{5/3}$ (upper axis) for a range of 1.7<T<17 K; (b) The field dependence of the coefficient $A_{ab}(\alpha=2)$ (left scale) and $A^*_{ab}(\alpha=5/3)$ (right scale) for B||ab-plane.

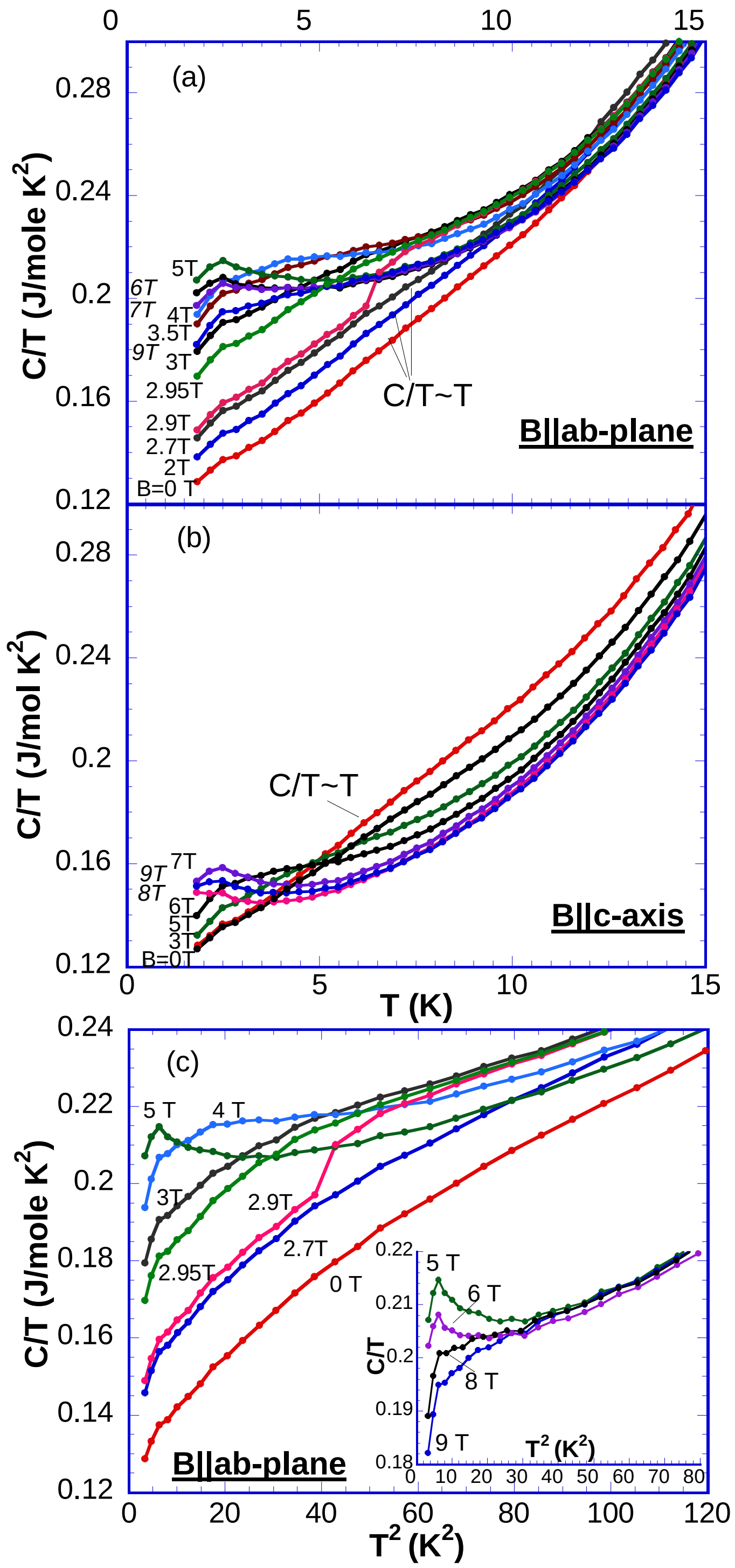

Fig.1

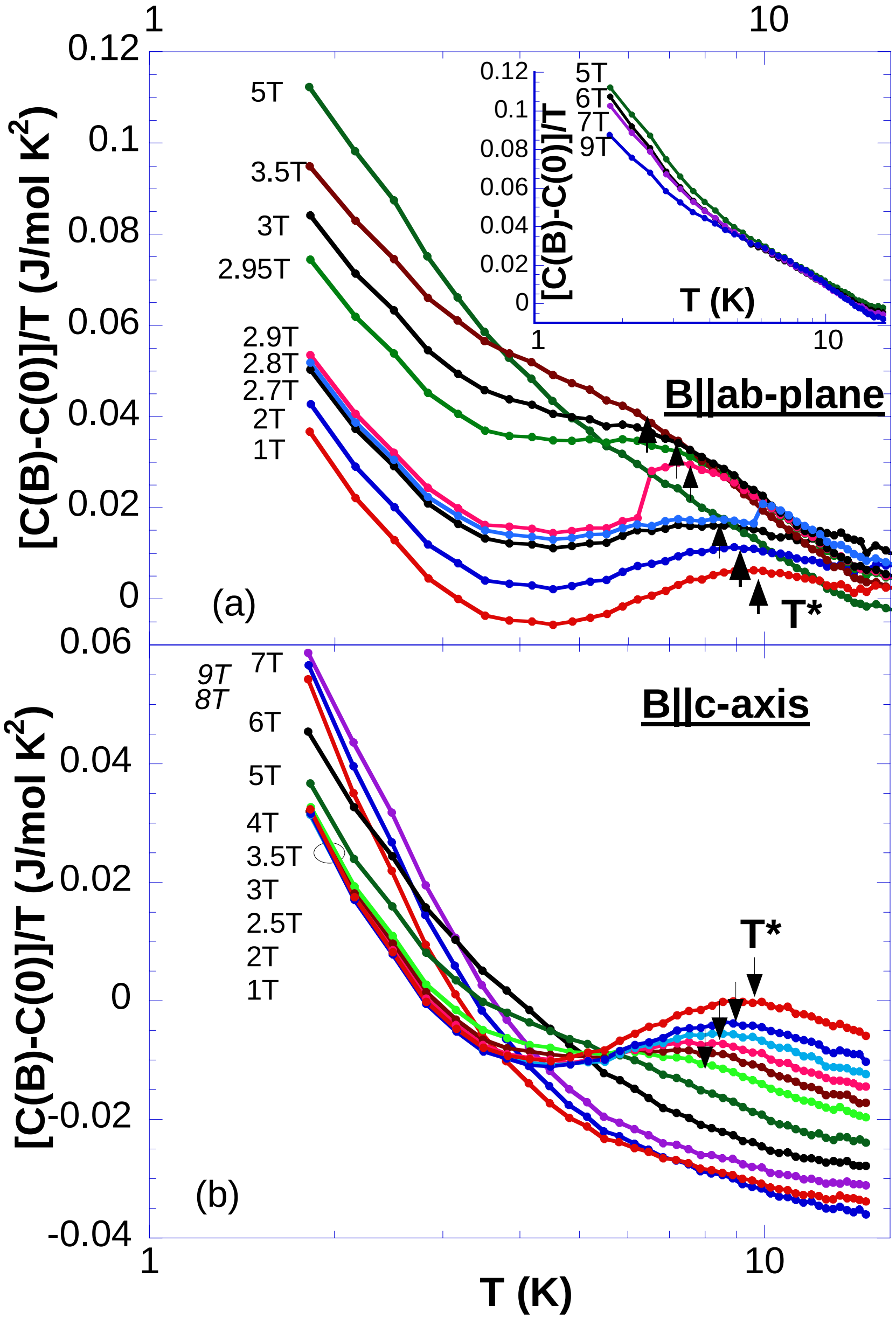

Fig.2

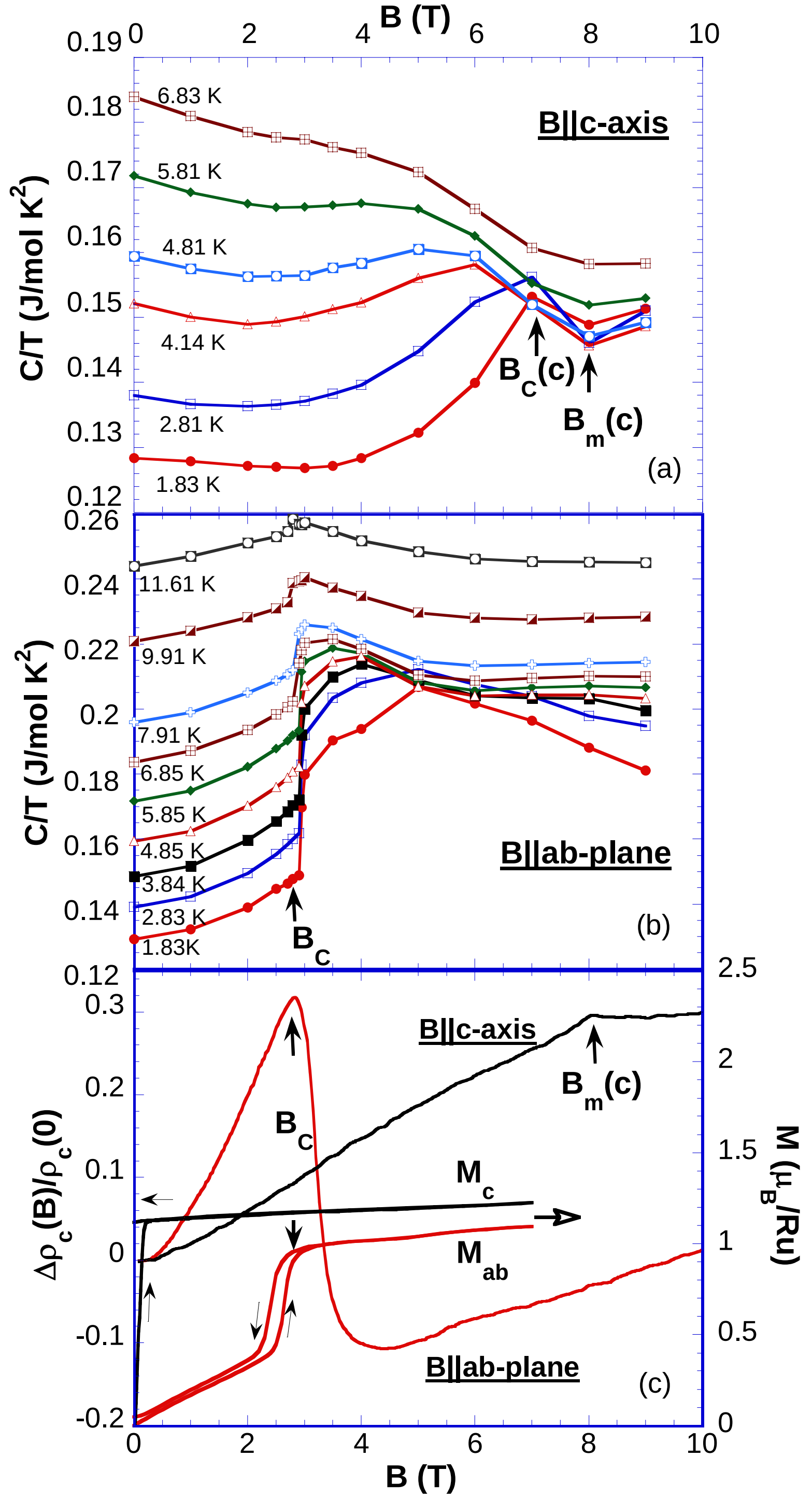

Fig. 3

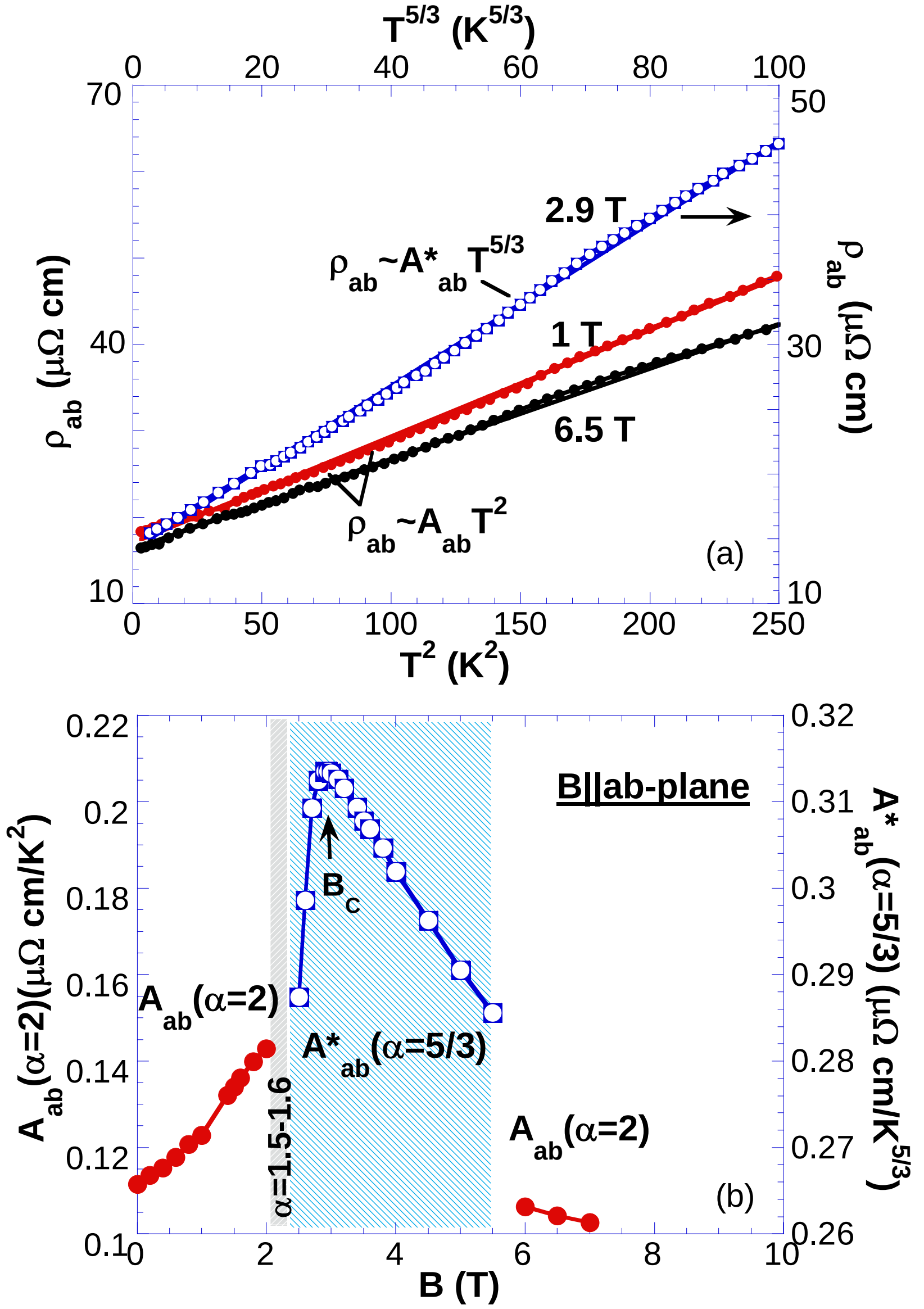

Fig. 4